\documentclass[12pt,preprint]{aastex}
\usepackage{amsmath}                
\usepackage{amsfonts}               
\usepackage{amssymb}                
\def \ev{~\rm{eV}}

\def \cm{~\rm{cm}}
\def \s{~\rm{s}}
\def \km{~\rm{km}}

\def \K{~\rm{K}}

\def \AU{~\rm{AU}}

\def \yr{~\rm{yr}}

\def \kpc{~\rm{kpc}}

\begin{document}

\title{MODELLING THE RADIO LIGHT CURVE OF ETA CARINAE}

\author{Amit Kashi\altaffilmark{1} and Noam Soker\altaffilmark{1}}

\altaffiltext{1}{Department of Physics, Technion$-$Israel
Institute of Technology, Haifa 32000 Israel;
kashia@physics.technion.ac.il;
soker@physics.technion.ac.il.}

\begin{abstract}

We study the propagation of the ionizing radiation emitted by the secondary star
in $\eta$ Carinae. We find that a large fraction of this radiation is absorbed by
the primary stellar wind, mainly after it encounters the secondary wind and passes
through a shock wave.
The amount of absorption depends on the compression factor of the primary wind in the
shock wave.
We build a model where the compression factor is limited by the magnetic pressure
in the primary wind.
We find that the variation of the absorption by the post-shock primary wind with orbital
phase changes the ionization structure of the circumbinary gas
and can account for the radio light curve of $\eta$ Car.
Fast variations near periastron passage are attributed to the onset of the accretion
phase.
\end{abstract}

\section{INTRODUCTION}
\label{sec:intro}

The $5.54 \yr$ light periodicity of the massive binary system $\eta$ Car
is observed from the radio (Duncan \& White 2003; White et al. 2005)
to the X-ray band (Corcoran 2005).
According to most models, the periodicity follows the 5.54~years periodic
change in the orbital separation in this highly eccentric,
$e \simeq 0.9$, binary system (e.g., Hillier et al. 2006).
The X-ray cycle presents a deep minimum lasting $\sim 70$~day and occurring
more or less simultaneously with the spectroscopic event
(e.g., Damineli et al.\ 2000), defined by the fading, or even disappearance,
of high-ionization emission lines  (e.g., Damineli 1996; Damineli et al.\ 1998, 2000
Zanella et al. 1984).
The spectroscopic event includes changes in the continuum and lines
(e.g., Martin et al. 2006,a,b; Davidson et al. 2005).
The X-ray minimum and spectroscopic event are assumed to occur near
periastron passages.

The radio emission comes from thermal bremsstrahlung (free free) emission
from ionized material (Cox et al. 1995; Duncan \& White 2003).
We follow previous suggestions that most of the UV radiation that ionizes
material at large distances from the binary system originates in the hot
secondary star (Verner et al. 2005; Duncan \& White 2003; Soker 2007a).
Duncan \& White (2003) further suggested that most of the radio emission
comes from a disk of a few arcsec, $\sim 10^4 \AU$ at the distance of $\eta$ Car,
around the binary system.
We accept the general picture suggested by Duncan \& White (2003) and
White et al. (2005), and try to explain the general properties of the radio emission.

The $3.5 \cm$-radio emission after the 1998 minimum was lower than in the previous
cycle (Duncan \& White 2003; White et al. 2005).
Duncan \& White (2003) attributed this fading to a denser equatorial gas due
to shell ejection at the 1998 periastron passage.
The shell ejection idea goes back to Zanella et al. (1984).
However, the shell ejection model fails to explain the properties of the X-ray
minimum (Akashi et al. 2006; see discussion in Soker 2007a), as well
as the behavior of some visible emission lines (Nielsen et al. 2007).
We will therefore examine other explanations, such as the secondary became
cooler and changes in the magnetic properties of the primary wind.

Our main goal is to understand the basic processes behind the variations in
radio luminosity in order to learn on the interaction of the two winds,
and to reveal signatures of accretion during the spectroscopic event.
In addition, as will be evident (see also Soker 2007a) the general flow structure
and ionization and recombination rates we deal with are far too sensitive to the
stellar and wind parameters to be modelled by a simple approach.
Therefore, we will not try to fit each point in the radio and millimeter light curves,
but only to approximately match the general behavior within a cycle, with the goal
of finding the role of the primary wind and the role of the secondary ionizing radiation.
We will also speculate on the reasons for the fast variations in the
intensity of the millimeter emission near periastron (Abraham et al. 2005).

\section{SETTING THE FLOW STRUCTURE}
\label{sec:flows}
\subsection{The Binary System}
\label{binary}
The $\eta$ Car binary parameters used by us are as in the previous
papers in this series (Soker 2005b; Akashi et al.\ 2006; Soker \&
Behar 2006; Soker 2007a), and are compiled by using results from
several different papers
(e.g., Ishibashi et al. 1999; Damineli et al. 2000; Corcoran et
al. 2001, 2004; Hillier et al. 2001; Pittard \& Corcoran 2002;
Smith et al. 2004; Verner et al. 2005).
The assumed stellar masses are $M_1=120
M_\odot$, $M_2=30 M_\odot$, the eccentricity is $e=0.9$, and
orbital period 2024 days, hence the semi-major axis is $a=16.64
\AU$, and the orbital separation at periastron is $r=1.66 \AU$.
The mass loss rates are $\dot M_1=3 \times 10^{-4} M_\odot
\yr^{-1}$ and $\dot M_2 =10^{-5} M_\odot \yr^{-1}$. The terminal
wind speeds are taken to be $v_1=500 \km \s^{-1}$ and $v_2=3000
\km \s^{-1}$.

\begin{table}

Table 1: Binary Parameters

\bigskip
\begin{tabular}{|l|l|c|}
\hline
Symbol  & Meaning   & Typical value \\
\hline \hline
$r$ & Binary separation& $1.66-31.6 \AU$  \\
\hline
$e$ & Eccentricity & 0.9     \\
\hline
$M_1$& Mass of primary   & $120 M_\odot$ \\
\hline
$M_2$& Mass of secondary   & $30 M_\odot$ \\
\hline
$\dot M_1$& Mass loss rate by primary   & $3 \times 10^{-4} M_\odot \yr^{-1}$ \\
\hline
$\dot M_2$& Mass loss rate by secondary & $10^{-5} M_\odot \yr^{-1}$ \\
\hline
$v_1$ & Velocity of primary wind (relative to primary)  &  $500 \km \s^{-1}$ \\
\hline
$v_2$ & Velocity of secondary wind (relative to secondary) & $3000 \km \s^{-1}$ \\
\hline
${v_{\rm wind1}}$ & Velocity of primary wind relative to secondary  &  $\sim 500 \km \s^{-1}$ \\
\hline
$R_2$ & Stellar radius of secondary& $\sim 0.1 \AU$  \\
\hline
$L_2$ & Secondary Luminosity   & $ 9\times 10^{5} L_\odot $  \\
\hline
$T_2$ & Secondary effective temperature & $36,000-40,000 \K$  \\
\hline
\end{tabular}
\footnotesize
\bigskip
 \normalsize
\end{table}

\begin{table}

Table 2: Geometrical and Other Parameters

\bigskip
\hskip -1.2 cm
\begin{tabular}{|l|l|l|c|}
\hline
Symbol  & Meaning   & Typical value & Defined \\
\hline \hline
$D_1$& Distance of stagnation point from primary
$D_1=r-D_{g2}$ & $0.7 r$ & Fig. 1 \\
\hline $D_{g2}$ & Distance of stagnation point from secondary & $0.3 r$ & Fig. 1 \\
\hline
$\phi_a$     & Asymptotic angle of contact discontinuity      & $60^\circ$  & Fig. 1      \\
\hline
$\phi$ & Angle from primary to a point on contact discontinuity   &$0-60^\circ$     & Fig. 2      \\
\hline
$\alpha$ & Angle from secondary to a point on contact discontinuity & $0-180^\circ$    & Fig. 2      \\
\hline
$\psi$  & Angle between primary wind and contact discontinuity    &  $0-90^\circ$   & Fig. 2     \\
\hline
$\delta$  & Angle between secondary wind and contact discontinuity    &  $0-90^\circ$   & Fig. 2     \\
\hline
$\theta$  & Orbital angle ($\theta=0$ at periastron) & $-180^\circ-180^\circ$ &       \\
 \hline
$t$ & Orbital phase     &  $0-1$ $(0-2024$~day)    &      \\
\hline
$r_{1s}$ & Distance from primary to contact discontinuity     &  $\sim 1-$~hundreds~AU   & Fig. 2      \\
\hline
$r_{2s}$ & Distance from secondary to contact discontinuity     &  $\sim 1-$~hundreds~AU   & Fig. 2      \\
\hline
$r_{20}$  & Inner boundary of recombining undisturbed primary wind      & $\ga 300 \AU$    &  eq. (\ref{rh1})   \\
 \hline
$r_{2i}$ &  Outer boundary of recombining undisturbed primary wind    & $\sim r_2s$ &  eq. (\ref{rh1})     \\
 \hline
$r_2$ & Distance from secondary  &   $0.5-$~hundreds~AU   &      \\
\hline
$d_p$ & Length a ray from secondary cut in the thin shell      & $\sim 0.2-$~hundreds~AU  &   Fig. 2    \\
 \hline
$v_d$ & Post-shocked primary wind speed in the thin shell     & $0-250 \km \s^{-1}$  &  Fig. 2   \\
 \hline
$q_v$ & Parameter determining the value of $v_d$ by eq. (\ref{vd})   &  0.5 &   eq. (\ref{vd})    \\
\hline
$\rho_1$ & Primary wind density     &  $\dot M_1/4 \pi v_1 r_1^2$   &   eq. (\ref{rho})    \\
\hline
$\rho_p$ & Density of post shock primary wind in the conical shell & $(1 - 40)\times \rho1$ & eq. (\ref{shockB})   \\
 \hline
$\eta_B$ & Preshock magnetic pressure to ram pressure ratio  & $0.001-0.1$    &   eq. (\ref{etaB})   \\
 \hline
$\dot R$ &Recombination rate in the undisturbed primary wind & $\sim 10^{48}-10^{50}{\rm sr}^{-1} \s^{-1}$& eq. (\ref{rh1})    \\
 \hline
$\dot R_{\rm shock}$ &Recombination rate in the conical shell & $\sim 10^{47}-2 \times 10^{49} {\rm sr}^{-1} \s^{-1}$& eq. (\ref{rec2})    \\
 \hline
$n_{p}$ & Proton density  & $10^5-10^{12} \cm^{-3}$    &      \\
 \hline
$n_e$ & electron density  &   $10^5-10^{12} \cm^{-3}$   &       \\
 \hline
$N_{i2}$ & Rate of emission of ionizing photons by the secondary& $2-3.5 \times 10^{48} \s^{-1}$ &  eq. (\ref{ni2}) \\
 \hline
$k_r$ & Efficiency of radio emission  &   $0.01-0.1$  & eq. \ref{lradio}     \\
\hline
\end{tabular}
\footnotesize
\bigskip
\normalsize
\end{table}

The dependence of the primary wind properties on latitude (Smith et al. 2003)
is not important here, because we are interested in the interaction winds
region which occurs near the equatorial plane. High latitudes are less
relevant to our goal.
The properties of the primary wind segments that interact with the secondary
wind are mainly constrained by the X-ray emission
(Pittard \& Corcoran 2002; Akashi et al. 2006), and are used here.
The general flow structure is depicted in Figure 1.
\begin{figure}
\resizebox{0.90\textwidth}{!}{\includegraphics{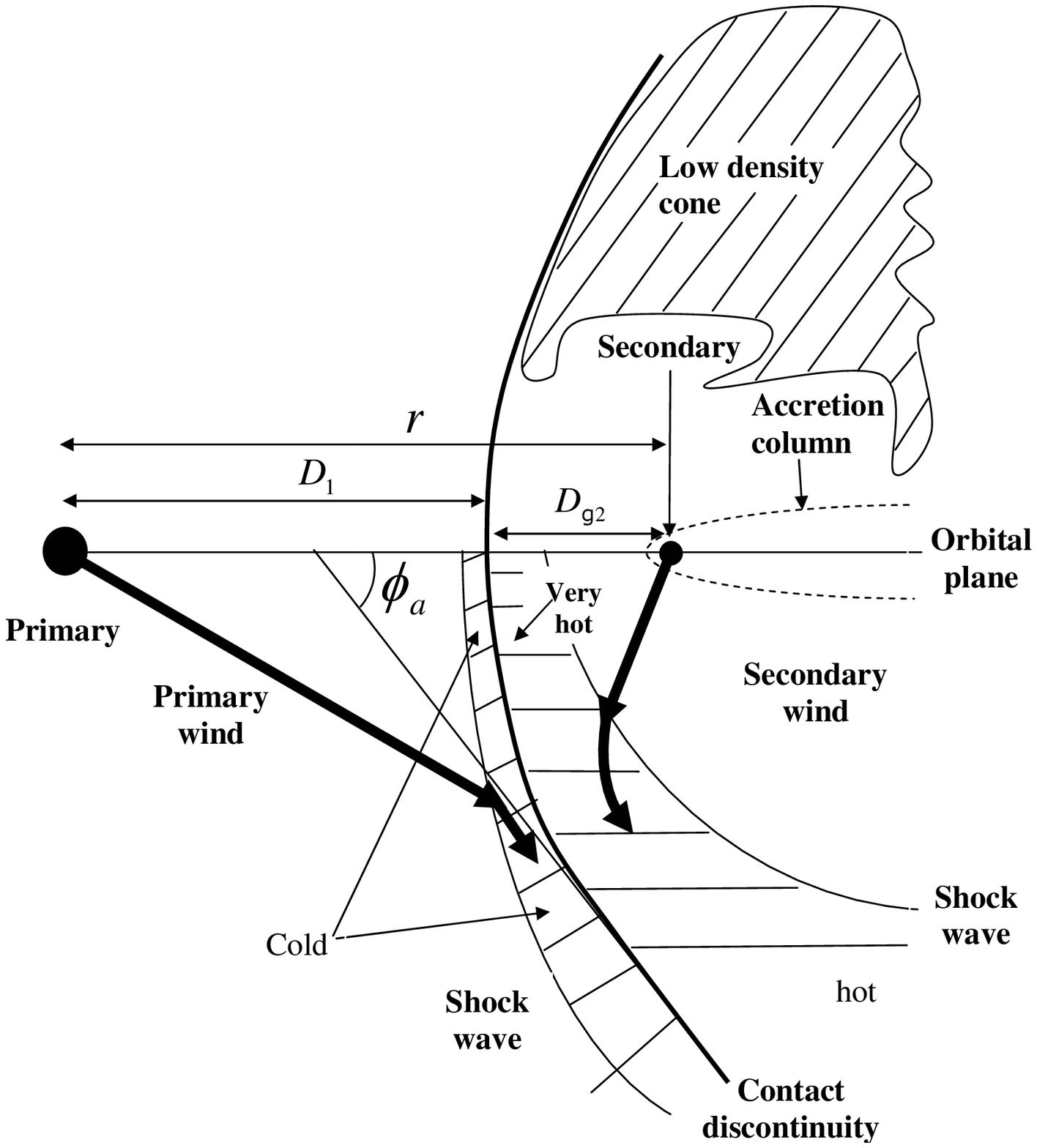}}
\vskip -5.2 cm
\caption{Schematic drawing of the collision region of the two stellar
winds and the definition of several quantities used in the paper.
There is an axial symmetry around the line through the two stars.
The two thick lines represent winds' stream lines.
The two shock waves are drawn only in the lower half.
The post-shock regions of the two winds are hatched.
The region dominated by the low-density secondary wind, to
the right of the contact discontinuity in the figure, is termed the
low-density cone (or shock cone).
The dashed line shows the accretion column which exists,
according to the proposed model, only for $\sim 70-80$~days
during the accretion period which corresponds to the X-ray  minimum
and the spectroscopic event.
} \label{fwinds1}
\end{figure}

Verner et al. (2005) deduced the following secondary stellar properties:
$T_2 \simeq 37,200 \K$, $L_2 \simeq 9.3 \times 10^5 L_\odot$,
$R_2 \simeq 23.6 R_\odot$, $v_2=2000 \km \s^{-1}$, and
$\dot M_2 \simeq 8.5 \times 10^{-6} M_\odot \yr^{-1}$.
The properties of the secondary star used by us are discussed in
previous papers of the series (Soker \& Behar 2006; Soker 2007a),
and are based on the results of Verner et al. (2005).
We take $R_2 \simeq 20 R_\odot$ and $T_2 \simeq 36,000-40,000 \K$.

Using the results of Schaerer \& de Koter (1997) for ionizing flux from hot stars,
we fit the rate of ionizing photons ($h\nu >13.6 \ev$) emitted by the secondary
star per steradian by
\begin{equation}
 \dot N_{i2} =\frac {\dot N_{i2-t}}{4 \pi} =
 \frac{L_2}{9\times 10^{5} L_\odot} \s^{-1} {\rm sr}^{-1}\left\{
\begin{array}{ll}
   3.5 \times 10^{48}    \quad T_2=40,000 \K  \\ 
   2.3 \times 10^{48}    \quad T_2=36,000 \K  \\ 
\end{array}\right.
\label{ni2}
\end{equation}
where a linear fit can be done for any effective temperature in the above range
(see Soker 2007a).

\subsection{Undisturbed Primary Wind}
\label{undisturbed}
\begin{figure}
\resizebox{0.99\textwidth}{!}{\includegraphics{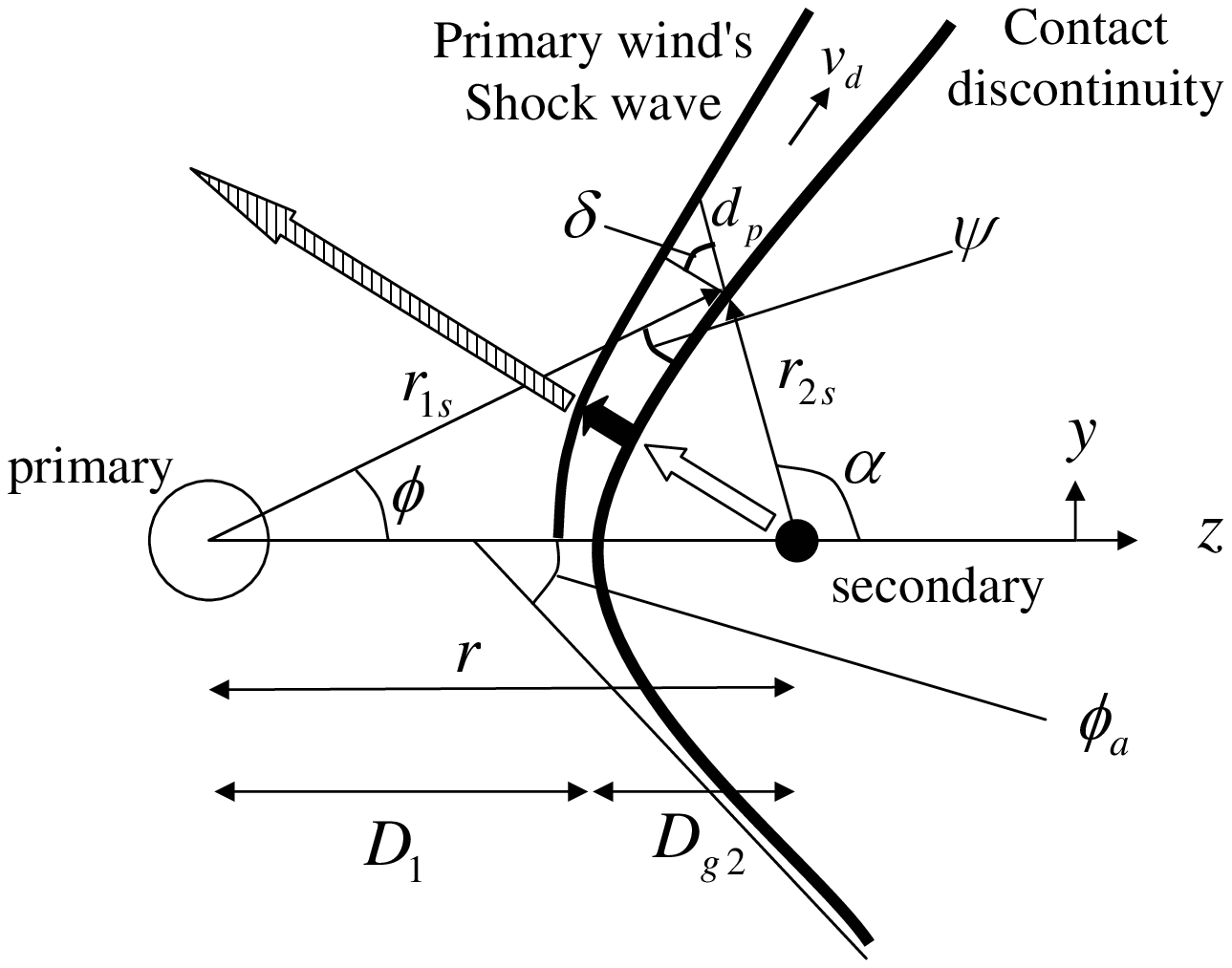}}
\vskip -9.2 cm
\caption{Schematic drawing of the collision region of the two stellar
winds and the definition of several quantities used in the paper.
The contact discontinuity is approximated as having hyperbolic shape.
The wide arrows represent absorption of ionizing photons:
The empty arrow is in the secondary wind region (low-density cone), where absorption
is negligible. The short black arrow is in the primary wind postshock
region, where most absorption occurs. The hatched arrow represent absorption
in the freely expanding primary wind region.
The figure is not to scale. For the parameters used here $D_{g2} \simeq 0.3 r$, and
$d_p \ll r_{2s}$ in the relevant regions.}
\label{fwinds2}
\end{figure}

We examine the fraction of the ionizing photons absorbed by the
undisturbed (freely expanding) primary wind.
Namely, taking the region from $r_2=r_{20}$ to
$r_2=r_{2i}$ to be the freely expanding primary wind, i.e.,
the secondary stellar wind and gravity do not deflect the primary
wind in that region.
Both $r_{20}$ and $r_{2i}$ can depend on the angle $\alpha$.
Inside the low-density cone the density of the secondary wind is very low, and it can be
neglected when calculating the recombination rate.
The different variables that will be used are defined in Figures
\ref{fwinds1} and \ref{fwinds2}.

The density of the primary wind as function of distance $r_2$ from the secondary is
\begin{equation}
\rho_1(r_2)=\frac{\dot M_1} {4 \pi v_1 (r^2+r_2^2+2 r_2 r \cos \alpha)},
\label{rho}
\end{equation}
where $r$ is the orbital separation and $\alpha$ the angle between the direction
of $r_2$ (measured from the secondary) and the line joining the two stars.
The total recombination rate per steradian along that direction is
\begin{equation}
\dot R(r_{20},r_{2i},\alpha) =\alpha_B \int_{r_{20}}^{r_{2i}} n_e n_p r_2^2 dr_2
= G(\dot M_1,v_1)
 \int_{r_{20}}^{r_{2i}} \frac{r_2^2}{(r^2+r_2^2+2 r_2 r \cos \alpha)^{2}} dr_2,
\label{rh1}
\end{equation}
where $\alpha_B$ is the recombination coefficient, $\mu m_H$ is the mean mass per
particle in a fully ionized gas
(we assume solar abundance), and
\begin{equation}
G(\dot M_1,v_1)=
0.22  \alpha_B  \left( \frac{\dot M_1}{4 \pi v_1 \mu m_H} \right)^2.
\label{g1}
\end{equation}

The integral in equation (\ref{rh1}) has an analytical solution that gives for
the recombination rate per steradian
\begin {equation}
\frac{\dot R(r_{20},r_{2i},\alpha)}{G(\dot M_1,v_1)}= \left[
\frac{2 r_{2} \cos^2 \alpha -r_{2} + r \cos \alpha}{2 \sin^2 \alpha
(r^2+r_{2}^2+2 r_{2} r \cos \alpha)} + \frac{1}{2 r \sin^3 \alpha}
\tan^{-1} \frac{r_{2}+r \cos \alpha}{r \sin \alpha} \right] _{r_{20}}^{r_{2i}} .
\label{rh2}
\end{equation}
Taking $r_{2i} \rightarrow \infty$ and $r_{20}=0$ gives the absorption
in the case where the primary wind is not disturbed at all
\begin{equation}
\dot R(0,\infty,\alpha) = G(\dot M_1,v_1)
\left(\frac{\alpha}{2 r \sin^3 \alpha} -\frac{\cos \alpha}{2 r \sin^2 \alpha} \right)
\label{rh3}.
\end{equation}
Substituting typical values for $G(\dot M_1,v_1)$ in equation (\ref{rh3}) we find
\begin{equation}
\dot R(0,\infty,\alpha) =  3.2 \times10^{47}   
\left(\frac{\dot M_1}{3 \times 10^{-4} M_\odot \yr^{-1}} \right)^2
\left(\frac{v_1}{500 \km \s^{-1}} \right)^{-2}
\left(\frac{r}{5 \AU} \right)^{-1}
\left(\frac{\alpha}{ \sin^3 \alpha} -\frac{\cos \alpha}{\sin^2 \alpha} \right) \s^{-1}
\label{rh4}.
\end{equation}

We note that the freely expanding primary wind absorbs radiation mainly in the direction
$\alpha \ga 140^\circ$, namely in directions toward the primary star.
However, equation (\ref{rh2}) largely overestimates the absorption of the secondary
ionizing radiation in those directions because the primary itself ionizes the material
close to it.

\subsection{The Shocked Primary Wind}
\label{shocked}

\begin{figure}
\resizebox{0.89\textwidth}{!}{\includegraphics{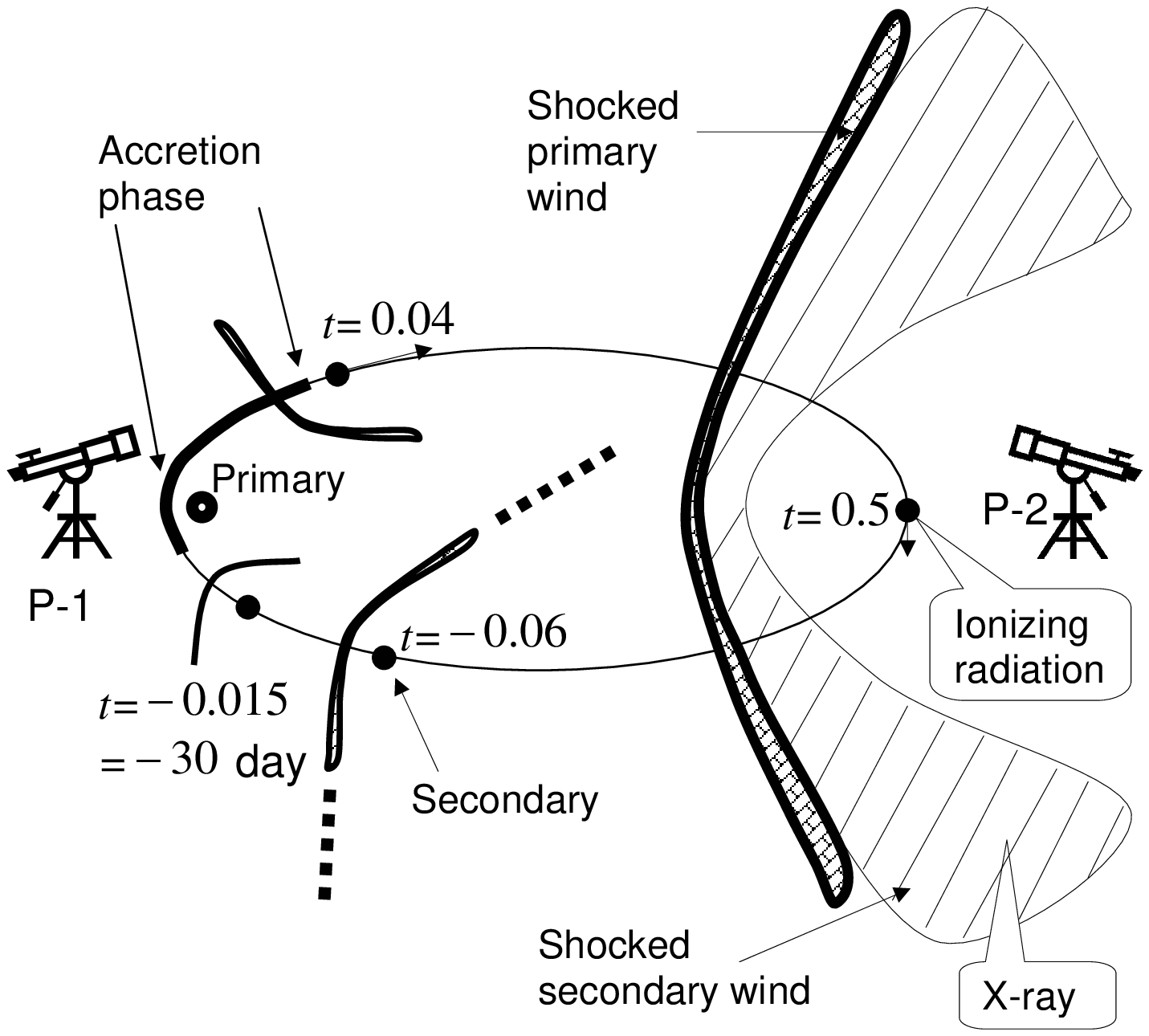}}
\vskip -7.2 cm
\caption{The orbital plane and a cut through the primary wind shocked region;
this region extends beyond the size of the figure, as drawn schematically at phase
$t=-0.06$.
The phases are marked at four points along the orbit.
The elliptical orbit, the primary stellar size, the distance of the shocked
primary wind region and its width, are drawn approximately to scale (but
not the secondary star or the secondary wind shock region).
The thick section of the orbit is approximately where accretion phase occurs according
to the model.
At phase $t=0.5$ we mark the sources of the X-ray emission (the shocked secondary wind), of the
ionizing radiation (the secondary photosphere).
Two possible orbital orientations are shown.
Position 1: The secondary is closer to us near periastron passage.
Position-2: The primary is closer to us near periastron
passage (Nielsen et al. 2007). In both cases the orbital plane is tilted
by $\sim 45 ^\circ$ to the line of sight. }
\label{orbitf}
\end{figure}

The secondary wind `cleans' the area behind the secondary (right side in Figure \ref{fwinds1})
and compresses the primary wind along the contact discontinuity, increasing the recombination
rate there.
For the parameters used here the half opening angle of the wind-collision
cone is $\phi_a \simeq 60 ^\circ$ (Akashi et al. 2006).
This implies that even as the secondary approaches or recedes periastron
the secondary fast wind will clean a large solid angle for ionization to propagate
almost unattenuated at low latitudes, unless the fast wind is shut down for $\sim 10$~weeks
as assumed here (Soker 2007a).
The general structure along the orbit is shown at 4 epochs in Figure \ref{orbitf}.

The fast orbital motion near periastron causes this low-density (shock) cone to wind (wrap)
around the secondary, and therefore to increase absorption (Nielsen et al. 2007).
This is important only very near to periastron passage, and it is much too complicated to deal
with in this analytical paper.
But most important, when this winding becomes important we expect an accretion phase
to start (Soker 2005b).
We will therefore, ignore winding of the wind interaction region around the binary system.

The primary wind postshock region is unstable, and has a corrugated structure
(Pittard \& Corcoran 2002; Pittard et al. 1998).
Magnetic fields within the primary wind will reduce the compression factor
as we shall see below.
Because of the unstable postshock flow structure and the unknown magnetic field
intensity, our calculation here is a crude estimate, but still teaches us
on the importance of the shock wave.
The post-shocked primary wind flows in a thin shell along the contact discontinuity
with a speed $v_d$, as drawn schematically in Figures \ref{fwinds1} and {\ref{fwinds2},
where all variables used here are defined.
Let the distance a ray from the secondary cuts in the thin shell be $d_p$, such that
the shell width (perpendicular to the flow) is $d_p \cos \delta$.

The contact discontinuity shape is approximated by an hyperbola at distance
$D_{g2}$ from the secondary at the stagnation point of the colliding winds,
where the two winds momenta balance each other along the symmetry axis,
and asymptotic angle $\phi_a=60^\circ$.
The velocity $v_d$ of the post-shock primary wind parallel to the shock front
is zero at the stagnation point, and increases to $\sim v_1$ at infinity.
Mixing between the two winds, as a result of instabilities  (the corrugated structure;
Pittard \& Corcoran 2002), will further accelerate the post-shock primary wind
(Girard \& Willson 1987).
This happens because the post-shock secondary wind expands faster than the
primary wind and because of its long cooling time it has a pressure gradient
parallel to the contact discontinuity.
Therefore, the value of $v_d$ is hard to calculate in the analytical model.
We therefore take this velocity to be
\begin{equation}
v_d=v_1 q_v \cos \psi,
\label{vd}
\end{equation}
where $q_v$ is a parameter that is kept constant in the model.
Increasing $q_v$ will have similar effect as reducing the density
compression ratio (see below). We take $q_v=0.5$.

The density of the postshock primary wind introduces large uncertainties,
because it is strongly influenced by the magnetic field.
The primary wind rapidly cools to a temperature of $T_p \simeq 10^4 \K$,
and it is compressed by the ram pressure of the slow wind to a density $\rho_p$.
Neglecting first the magnetic pressure in the post-shock region,
$\rho_p$ is given by equating the thermal pressure $kT_p \rho_p/\mu m_H$ of the post-shock material
with the ram pressure of the primary wind $\rho_1 (v_{\rm wind1} \sin \psi)^2$,
where $\psi$ is the angle between the slow wind speed and the shock front,
and $v_{\rm wind1}$ is the preshock speed of the primary wind relative to the stagnation
point;.
Because in most interesting orbital phases the orbital velocity is low,
we take here $v_{\rm wind1}=v_1$.
For a post-shock temperature of $T_p \simeq 10^4 \K$ we find for the compression factor
neglecting the magnetic field
\begin{equation}
\left( \frac{\rho_p}{\rho_1} \right)_{\rm thermal} \simeq 500      
\left( \frac{v_1}{500 \km \s^{-1}} \right)^{2}
\left( \frac{\sin \psi}{0.5} \right)^{2}.
\label{shock1}
\end{equation}
Such a high density contrast is seen in Figure 2 of Pittard \& Corcoran (2002).
However, because of magnetic fields we don't expect such a large compression
behind the shock.
Typical preshock magnetic pressure to ram pressure ratio can be
\begin{equation}
\eta_B \equiv P_{B0}/\rho_1 v_1^2 \sim 0.001-0.1
\label{etaB}
\end{equation}
(e.g., Eichler \& Usov 1993; Pittard \& Dougherty 2006).
The magnetic field component parallel to the shock is increased as it
is compressed when the density is increased in the shock wave.
For a random field we can take this component to contribute $1/3$ to the preshock pressure.
Equating the post-shock magnetic pressure to the wind's ram pressure we find the limit
on the compression factor imposed by the magnetic field to be
\begin{equation}
\left( \frac{\rho_p}{\rho_1} \right)_B = \left( \frac{3}{\eta_B} \right)^{1/2} \sin \psi
\simeq 19 
 \left( \frac{\eta_B}{0.002} \right)^{-1/2}  \frac{\sin \psi}{0.5} .
\label{shockB}
\end{equation}
The strong magnetic field can also smooth the strong corrugated structure seen in
the simulations presented by Pittard \& Corcoran (2002).
The presence of the magnetic field is another source of the large uncertainties
involved in our calculation, and it can introduce large stochastic variations
on short time scale and from cycle to cycle.
We will use equation (\ref{shockB}) to calculate the density $\rho_p$ because
$(\rho_p / \rho_1)_B < (\rho_p / \rho_1)_{\rm thermal}$ in the cases studied here.
In Figure \ref{etaf} we plot $\rho_p / \rho_1$ as function of the angle $\alpha$.
\begin{figure}
\resizebox{0.89\textwidth}{!}{\includegraphics{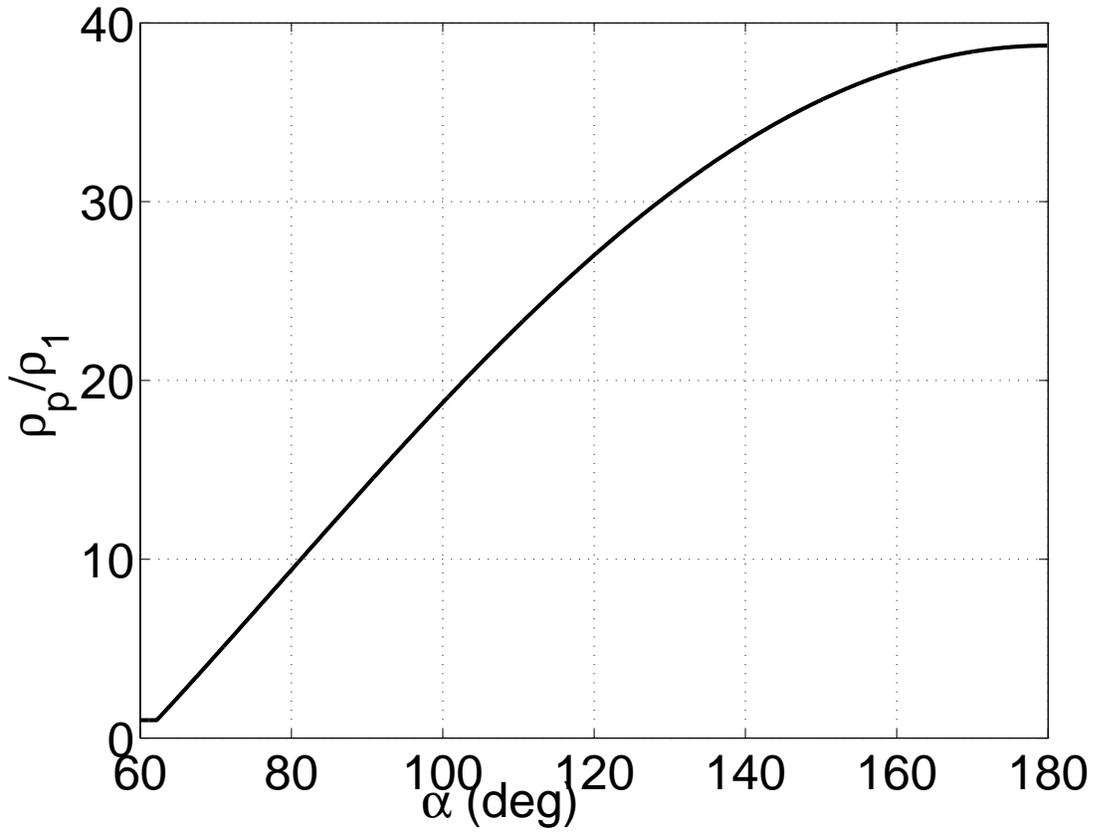}}
\caption{The density compression factor in the shock of the primary wind for
preshock magnetic pressure to ram pressure ratio of $\eta_B=0.002$.
In the entire range the density ratio is determined
by equation (\ref{shockB}), demonstrating the very important role of the
(weak) magnetic field in the preshock primary wind. If equation (\ref{shockB}) gives
${\rho_p}/{\rho_1}<1$ we set ${\rho_p}/{\rho_1}=1$ }
\label{etaf}
\end{figure}

Mass conservation for the primary wind implies that the mass expelled within an angle
$\phi$ equals the flow in the postshock shell.
This reads
\begin{equation}
\dot M_1 (\phi) \equiv \frac{1}{2} (1- \cos \phi)
= 2 \pi r_{2s} \sin \alpha v_d \rho_p \cos \delta d_p ,
\label{shock2}
\end{equation}
where $r_{2s}$ is the corresponding distance of the contact discontinuity from the
secondary.
The quantity $d_p$ is calculated from equation (\ref{shock2}).
The ratio $d_p \cos \delta /r_{2s}$ as function of $\alpha$ is plotted in Figure \ref{dpf}.
\begin{figure}
\resizebox{0.89\textwidth}{!}{\includegraphics{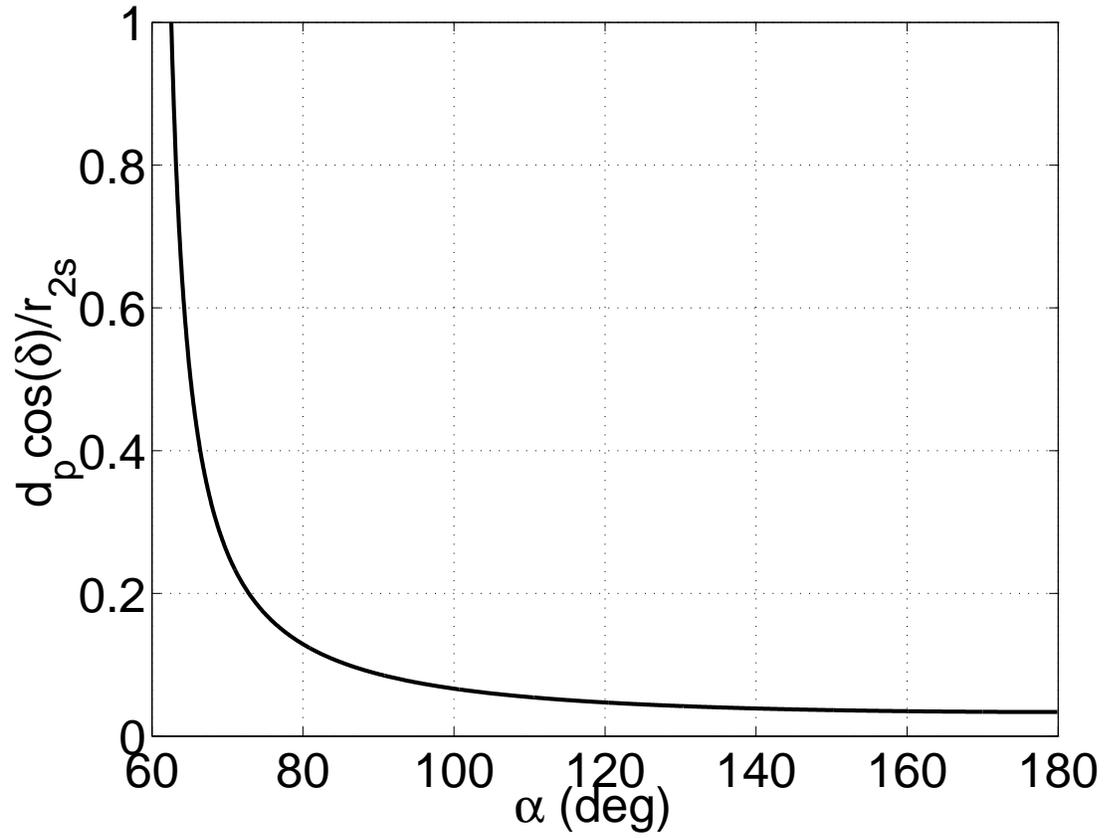}}
\caption{The ratio of the postshock shell width $d_p \cos \delta $ to $r_{2s}$
for the standard parameters listed in section \ref{results}
(see Figure \ref{fwinds2} for definitions of these quantities).}
\label{dpf}
\end{figure}

The recombination rate per steradian in the postshock shell is given by
\begin{equation}
\dot R_{\rm shock} = \alpha_B (n_e n_p)_p r_{2s}^2 d_p,
\label{rec2}
\end{equation}
where, as before, subscript `$p$' refers to postshock quantities
{{{  (e.g., $(n_e n_p)_p$ is the value of the electron number density times proton number density
calculated for the postshock gas). }}}
Using equation (\ref{g1}) we can scale equation (\ref{rec2}) and write
\begin{equation}
\dot R_{\rm shock} = 10 G(\dot M_1,v_1)
\left( \frac{\rho_p/\rho_1}{20} \right)^2
\left( \frac{r_{2s}}{0.5r_{1s}} \right)^2
\left( \frac{d_p}{0.1r_{1s}} \right)
\frac{1}{r_{1s}}.
\label{rec3}
\end{equation}

\subsection{The Radio Emission}
\label{radio}

The optical depth for radio emission (e.g., eq. 4.32 in Osterbrock 1989)
is given by
\begin{equation}
\tau \simeq 0.1    
\left( \frac{T}{10^4 \K} \right)^{-1.35}
\left( \frac{\lambda}{cm} \right)^{2.1}
\left( \frac{n_e}{10^6 \cm^{-3}} \right)^{2}
\left( \frac{h}{100 AU} \right)
\label{tau1}
\end{equation}
where $h$ is the typical size of the region considered, and
$n_e$ is the electron density.
For the undisturbed primary wind and fully ionized gas
\begin{equation}
n_e = 7.56 \times 10^5
\left( \frac{\dot M_1}{3 \times 10^{-4} M_\odot \yr^{-1}} \right)
\left( \frac{v_1}{500 \km \s^{-1}} \right)^{-1}
\left( \frac{r_1}{300 AU} \right)^{-2}
\cm^{-3} .
\label{ne1}
\end{equation}
We see from equation (\ref{tau1}) that for $\lambda \sim 3 \cm$ the optical
depth is $\tau > 1$ at distances  $r_1 \la 300 \AU$.

The contribution of the volume inner to the distance $r_1 \sim 300 \AU$,
where $\tau(3 \cm) \simeq 1$, is small because it is optically thick at these wavelengths.
We therefore take $r_{2i} =300 \AU$, which is $\sim 10$ times larger than apastron distance,
in equation (\ref{rh2}) when calculating absorption of ionizing photons by
the freely expanding primary wind. The lower limit in equation (\ref{rh2}) is
$r_{20}=r_{2s}$.

Most radio emission comes from much larger distances, up to $\sim 10^4 \AU$
(Duncan \& White 2003; White et al. 2005).
There we expect that the emission will come from dense clumps formed by
instability of the outflowing primary wind.
We therefore assume that many clumps within a distance of
$\sim 1^{\prime \prime} =2300 \AU$, and density as the wind
density is at $r_1 \sim 300 \AU$ from the primary star,
absorb a large fraction of the ionizing radiation escaping the binary system
and they are responsible for a large fraction of the radio emission.
They are optically thick at $\lambda \sim 3 \cm$.
In addition, some ionizing radiation escape to larger distances,
leading to no contribution to ionizing gas that emits in the radio band.
We therefore use an efficiency parameter $k_r (\lambda)$
to convert from maximum possible radio flux to observed radio flux.
The maximum possible radio emission is for the case where all ionizing radiation
ionizes optically thin gas to $\sim 10^4 \K$ in the observed region and in a steady state.
$k_r$ increases with decreasing wavelength with $k_r (3 \cm) \sim 0.1$.
Higher intensities are expected for shorter wavelength,
where optical depth is lower.
Indeed, emission flux at $\lambda=1.2 \cm$ is $\sim 2.5$
times as bright as that for $\lambda=3.5 \cm$ in 2004 (White et al. 2005).

We define the difference between ionizing photon rate emitted by the
secondary, $\dot N_{i2}$, and the total absorption
\begin{equation}
\dot N_{\rm diff}= \dot N_{i2}-[\dot R(r_{2s},300 \AU,\alpha)+\dot R_{\rm shock}(\alpha)].
\label{ndiff}
\end{equation}
This quantity can have negative values, which is meaningless.
In calculating the emitted radio luminosity
when ever $\dot N_{\rm diff} <0$ we set it to zero.
The radio emission at each orbital angle $\theta$ is calculate according to the
formula
\begin{equation}
L_{\rm radio} =k_r \Lambda 2 \pi
\int_{\alpha=0}^{180^\circ}
{\rm max} (\dot N_{\rm diff}, 0) \sin \alpha d \alpha,
\label{lradio}
\end{equation}
where $\Lambda$ is the conversion factor from ionizing photon rate (in equilibrium
with recombination) to radio power at $10^4 \K$ (Osterbrock 1989).

\section{RESULTS}
\label{results}

\subsection{The Basic Processes}
The parameters of the binary systems are given in section 2.1.
The other parameters for our standard case are as follows.
The contact discontinuity hyperbola has a distance
$D_{g2}=0.3 r$ from the stagnation point and an asymptotic angle $\phi_a=60^\circ$.
The velocity parameter in equation (\ref{vd}) is $q_v=0.5$,
the magnetic pressure in the primary wind $\eta_B=0.002$,
the radio efficiency $k_r=0.07$, and the ionizing photon rate per
steradian of the secondary star
$\dot N_{i2}=2\times 10^{48} \s^{-1} {\rm sr}^{-1}$  (eq. \ref{ni2}).
We could change the parameters in a way that will not change the results much.
For example, we can increase $\dot N_{i2}$ together with increasing a little the
compression ratio in the shock (by lowering $\eta_B$)
and reducing the radio efficiency.
These parameters are sufficient for our goal of demonstrating the importance of
absorption in the shocked primary wind, and connect this to the fast variation
in millimeter emission near periastron (section \ref{peri}).

First we examine the importance of the absorption by the primary wind of the
ionizing radiation emitted by the secondary star.
In Figure \ref{absf} we draw the absorption rate (assumed to be equal to the recombination rate)
per steradian as function of the angle $\alpha$ as defined in Figure \ref{fwinds2}:
$\alpha=0^\circ$ is away from the primary and $\alpha=180^\circ$ is toward the primary.
The thick lines in the left panel show the absorption due to the shocked
primary wind as given by equation (\ref{rec3}), and the thin lines show the
absorption of the freely expanding primary wind as given by equation (\ref{rh2}),
from the shock position $r_{20}=r_{2s}(\alpha)$ to a distance $r_{2i} = 300 \AU$.
In the right panel we show the value of  $\dot N_{\rm diff}$ (eq. \ref{ndiff}),
the difference between ionizing photon rate emitted by the
secondary, $\dot N_{\rm diff}$ (eq. \ref{ndiff}), for
$\dot N_{i2}=2 \times 10^{48} \s^{-1} {\rm sr}^{-1}$ (dashed line in the figure),
and the absorption rate.
\begin{figure}
\resizebox{0.49\textwidth}{!}{\includegraphics{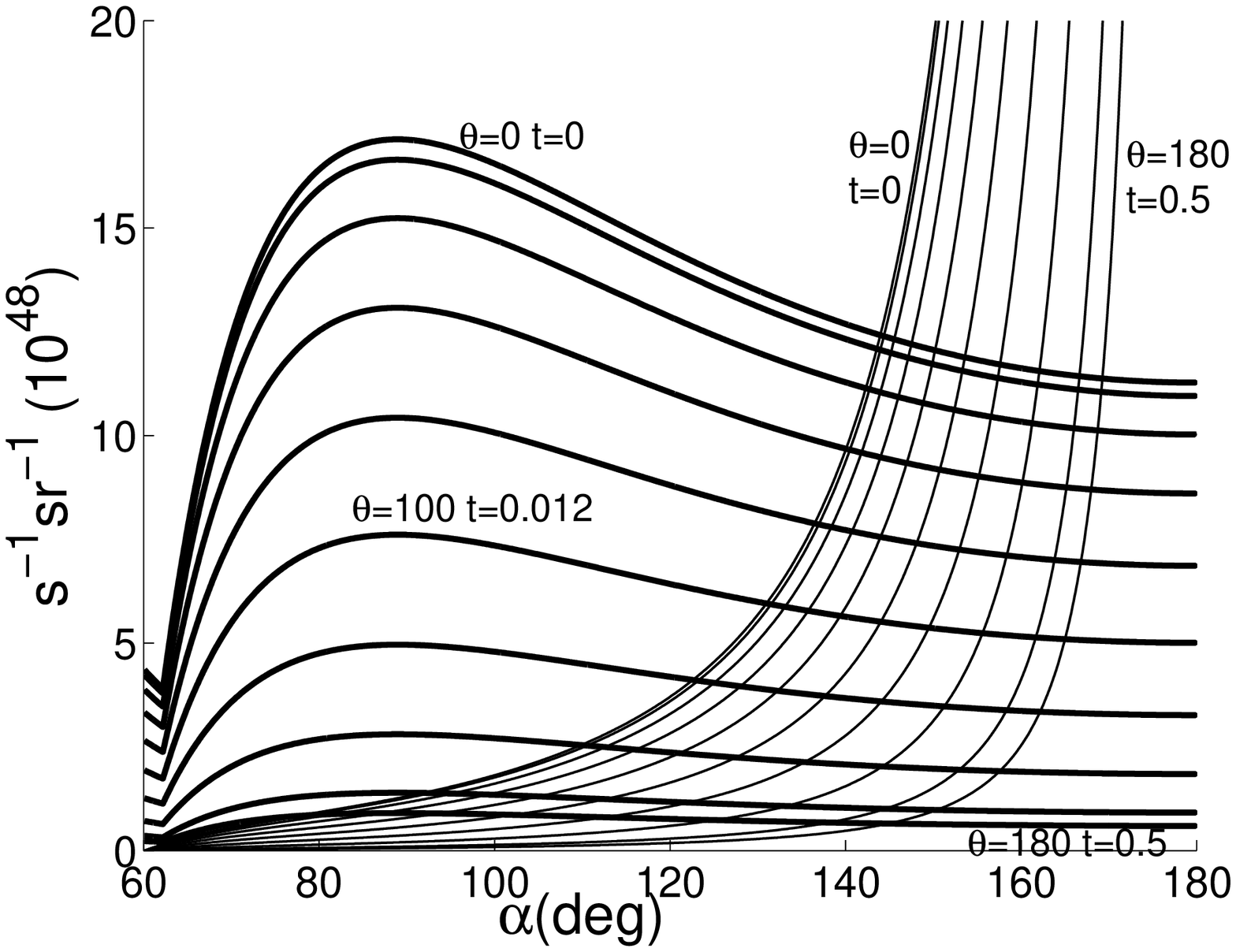}}
\resizebox{0.49\textwidth}{!}{\includegraphics{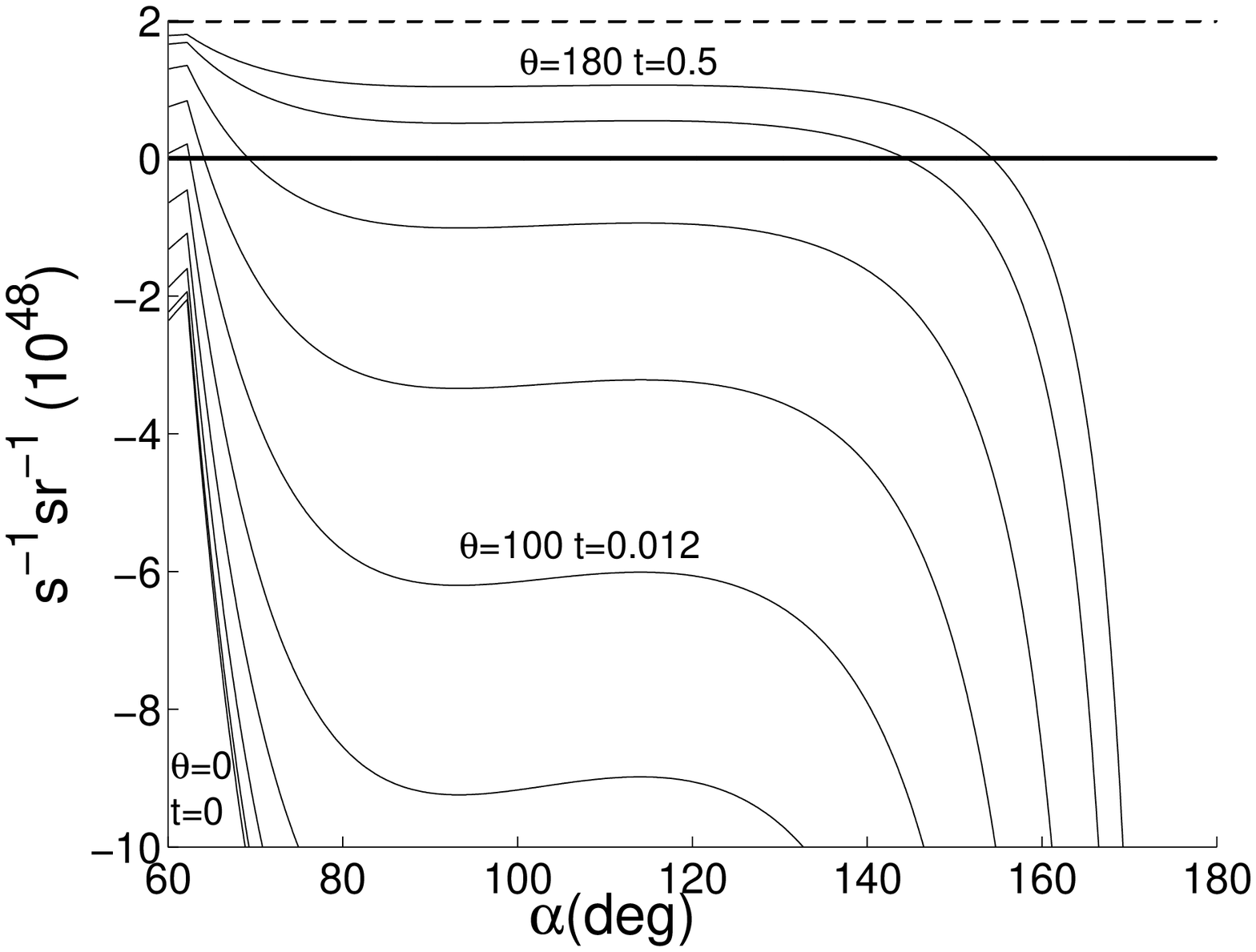}}
\caption{
Absorption rate and emission rate per steradian, for the standard parameters
given at the beginning of section 3, given every 20 degrees of the orbital angle
from $\theta=0$ (periastron) to $\theta=180^\circ$ (apastron).
Left panel: Absorption rate per steradian of ionizing photons by the primary wind,
assumed to be equal to the recombination rate.
The absorption rate of the shocked primary wind as given by equation
(\ref{rec3}) are drawn by thick lines.
The absorption rate of the undisturbed primary wind, as given by equation (\ref{rh2}),
from the shock position $r_{20}=r_{2s}(\alpha)$ to a distance $r_{2i} = 300 \AU$
are drawn by the thin lines.
Right panel: The difference between the secondary ionizing rate
$\dot N_{i2}=2 \times 10^{48} \s^{-1} {\rm sr}^{-1}$ (dashed line) and the total absorption,
as given by equation (\ref{ndiff}).
$t$ stands for the phase, with $t=0$ at periastron and $t=0.5$ at apastron.}
\label{absf}
\end{figure}

For $\alpha<60^\circ$ the secondary wind cleans a low-density cone, and no absorption
occurs there; this region is not in the graph.
(We don't consider the winding of this low-density cone around the primary near
periastron passage, which will increase absorption in directions  $\alpha<60^\circ$;
Nielsen et al. 2007.)
The absorption by the free wind is overestimated for large values of $\alpha$.
The reason is that these directions pass close to the primary star, where the primary
radiation ionizes a large volume around the primary. This is not considered
in our calculation.

The sharp change in the slope of the lines presenting the absorption by the shocked wind
near $\alpha=60^\circ$ result from our constraint that the density compression ratio cannot
be ${\rho_p}/{\rho_1}<1$, while close to the $\alpha=60^\circ$ equation (\ref{shockB}) gives
a value $<1$.
The same change in behavior is imprinted in the plots of $\dot N_{\rm diff}$ in
the right panel.
The results close to the asymptotic lines of the hyperbolic shock wave
are not trustable, but in any case this region plays a minor role.

In Figure \ref{radiof1} we show the radio light curve for our standard case
as calculated by equation (\ref{lradio}) with $k_r=0.07$.
We observe two orbital phases (and their two symmetric phases that are not shown)
where a change in behavior occurs in the light curve.
Near periastron, at phase$<0.011$ (time$<22$~day before or after periastron
passage) corresponding to orbital angle $\theta < 95^\circ$, the radio flux is constant.
This is the region where all ionizing photons in our model are absorbed by the
primary wind, beside those emitted into the low-density cone $\alpha <60^\circ$.
This region is not modelled correctly by us for two reasons.
First, the low-density cone is winding around the binary system (Nielsen et al. 2007)
as its orbital velocity is large ($\sim 300-400 \km \s^{-1}$) in this part of the orbit.
Second, and more important, at this phase an accretion process is likely to
start, where the secondary accretes from the primary wind (see section \ref{peri})
and the low-density cone does not exist for $\sim 10~$weeks.
\begin{figure}
\resizebox{0.89\textwidth}{!}{\includegraphics{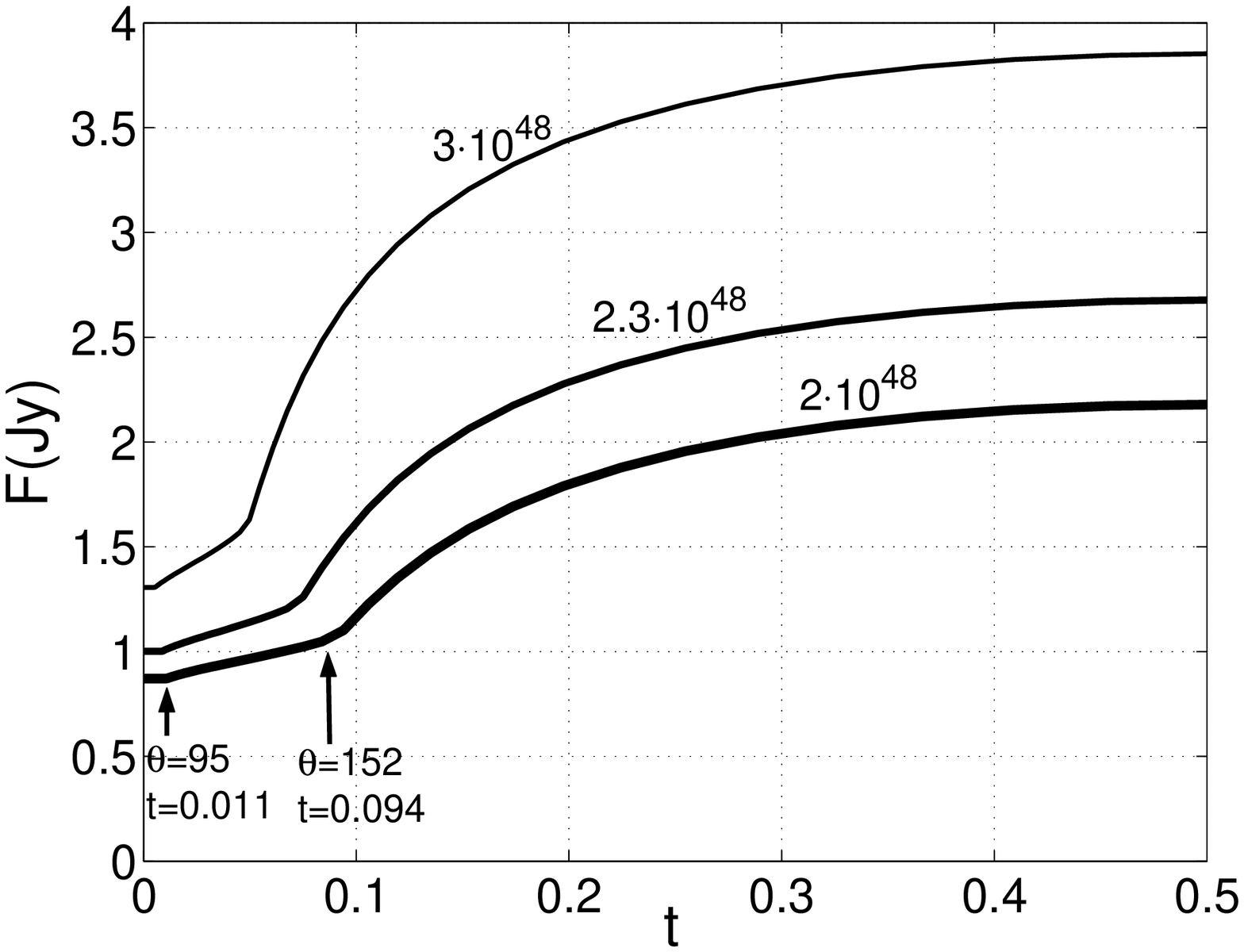}}
\caption{The calculated radio-flux light curve.
$F=L_{\rm radio}/ 4 \pi D^2$, where $L_{\rm radio} $ is given by
equation (\ref{lradio}) (with $k_r=0.07$), and $D=2.3 \kpc$ is the distance to $\eta$ Car.
The lower line is for the standard parameters
(given at the beginning of section 3).
The two other lines are for higher secondary ionizing photon rate,
as marked near each line in units of ionizing photons per second per steradian.
$t$ stands for the phase, with $t=0$ at periastron and $t=0.5$ at apastron.}
\label{radiof1}
\end{figure}

The second change in behavior occurs at phase$\simeq 0.094$
($\simeq 191$~day) corresponding to orbital angle $\simeq 152^\circ$.
As can be seen from the right panel of Figure \ref{absf}, this occurs when the
absorption in the directions $80^\circ \la \alpha \la 150^\circ$ drops below the
secondary ionizing photon flux.
For other parameters the change will occur at a different orbital phase.

\subsection{Periastron Passage}

According to the accretion model the deep X-ray minimum is assumed to result
from the collapse of the collision region of the two winds onto
the secondary star (Soker 2005b; Akashi et al. 2006).
This process is assumed to start $\sim 0-20$~day before periastron passage, to
last $\sim 10 ~$weeks, and shut down the secondary wind, hence the main X-ray source.
During that period the secondary accretes mass from the primary wind at a rate of
$\sim 10^{-6} M_\odot \yr^{-1}$.
Akashi et al. (2006) showed that this assumption provides a phenomenological
description of the X-ray behavior around the minimum.
We examine the implications of the accretion process to the model here and in section 4.

After the secondary wind is shut down there will be no low-density cone
behind the secondary any more.
The accretion process results in a high density region behind the secondary star,
the accretion column (dashed line in Figure \ref{fwinds1}).
No radiation will escape in that direction.
However, there is no shocked primary wind any more, and away from the equatorial
plane the absorption will be that of the freely expanding primary wind according
to equation (\ref{rh4}).
For our standard case we find that perpendicular to the line joining the two stars,
$\alpha=90^\circ$, and at apastron $r=1.66 \AU$, the absorption rate is
$\dot R(0,\infty,90^\circ) = 1.5 \times 10^{48} \s^{-1} {\rm sr}^{-1}$.
Since $\dot R(0,\infty,90^\circ)<N_{i2}$, we expect ionizing radiation to escape
to large distance. Therefore, the ionizing radiation is not expected to
drop to zero even during the $\sim 70$~day long accretion phase.

But even if ionizing radiation drops to zero, the radio emission will not follow.
The reason is that the recombination time (for our standard parameters)
\begin{equation}
\tau_{\rm rec} = \frac{n_p}{n_e n_p \alpha_B} =
60  
\left(\frac{r}{300 \AU} \right)^2
~{\rm day}
\label{trec}
\end{equation}
at relevant distances is longer than the accretion phase.
Many dense regions which give more radio emission will recombine
on a shorter time scale.
This holds along the entire orbit.
The recombination time at a distance of $\sim 1 ^{\prime \prime} \simeq 2300 \AU$
is longer than the orbital period,
and must be considered in trying to exactly fit the light curve.
But this effect is much more important near periastron passages, when changes
are fast and the wind segments as close as few hundred AU from the system
don't have time to recombine.

To summarize this section, it is possible that when the system is near periastron there
is still some ionizing flux reaching large distances.
But even if not, the accretion phase last only $\sim 10~$weeks, and the relevant regions
of the wind have no time to fully recombine.

\subsection{The Role of the Different Variables}

Because the ionizing radiation that escape to large distances is the difference between
the emitted and absorbed ionizing radiation, both the radio emission
(section 3.1) and the highly ionized emission lines in the UV and visible bands
(Soker 2007a) are very sensitive to the parameters of the system.

The ionizing radiation rate $\dot N_{i2}$ has an obvious influence.
In Figure \ref{radiof1} we present the results for increasing $\dot N_{i2}$ by a factor
of 1.15 to $\dot N_{i2}=2.3 \times 10^{48} \s^{-1} {\rm sr}^{-1}$,
and by a factor of 1.5 to $\dot N_{i2}=3 \times 10^{48} \s^{-1} {\rm sr}^{-1}$.
We see how the resulting radio emission at maximum increases by a larger factor
than the increase in $\dot N_{i2}$.
Near periastron the radio emission results from region ionizes by the
ionizing photons escaping only from the low-density cone, and hence it
increases by the same factor as $\dot N_{i2}$ was increased.
As discussed above, for $\sim 10$~week  very close to periastron passage an
accretion process is likely to occur, and the low-density cone does not exist.

The role of the primary mass loss rate and velocity is clearly evident from
the role of $G(\dot M_1,v_1) \propto \dot M_1^2/v_1^2$ in equations
(\ref{rh1})- (\ref{rh3}) and (\ref{rec3}). The absorption rate is sensitive to these
parameters, and therefore the ionizing radiation escaping to large distances.
Changing these parameters not only affect the absorption by the wind, but
also change the opening angle of the low-density cone.
For the parameters used here the dependance is approximately (Eichler \& Usov, 1993)
$\phi_a \propto (\dot M_2 v_2/\dot M_1 v_1)^{1/3}$.
The radiation escaping through the cone is a fraction $(1-\cos \phi_a)/2$
of the total secondary ionizing radiation.
The sense of the changes in the radio luminosity is opposite to that of
the primary mass loss rate, i.e., a higher value of $\dot M_1$ results
in a lower radio luminosity.

The primary wind speed, on the other hand, might have an opposite role near apastron
and periastron.
Say the primary wind decreases by a factor of $\sim 1.5$ to $v_1 \simeq 330 \km \s^{-1}$.
This implies a higher density in the wind.
For our standard case we find that for the entire orbit the total absorption is
above the emission rate, such that the only ionizing radiation escape from the
low-density cone.
Using the formula for the low-density cone opening angle as given by Eichler \& Usov (1993)
we find that it increases from $\phi_a=61 ^\circ$ to $\phi_a=68 ^\circ$.
The fraction of secondary ionizing photons that escape to large distances
increases from $\sim 0.26$ to $0.32$.
Remembering that near periastron most of the ionizing radiation escapes from the cone
(but not during the $\sim 10$~weeks of the accretion phase very close to periastron
passage, when the low-density cone does not exist),
we expect an increase of radio emission by a factor $\sim 20 \%$ near periastron passage,
but a decrease in radio emission near apastron, compared to our standard case.

Another important parameter is the density compression ratio in the shock.
Reducing the magnetic pressure $\eta_B$ in the preshock wind will allow
larger compression, i.e., larger  ${\rho_p}/{\rho_1}$ in equation (\ref{shockB}).
This results in higher absorption and lower observed radio flux.
In Figure \ref{compare} we show our standard case plotted for the 1992-8 cycle,
and a case of $\eta_B=0.001$, with all other parameters unchanged,
plotted on the 1998-2003 cycle.
The observed radio light curve is from White et al. (2005).
We don't try to fully fit the radio observations, but just to show the effect
of varying the value $\eta_B$, and by that emphasizing its importance.
\begin{figure}
\resizebox{0.99\textwidth}{!}{\includegraphics{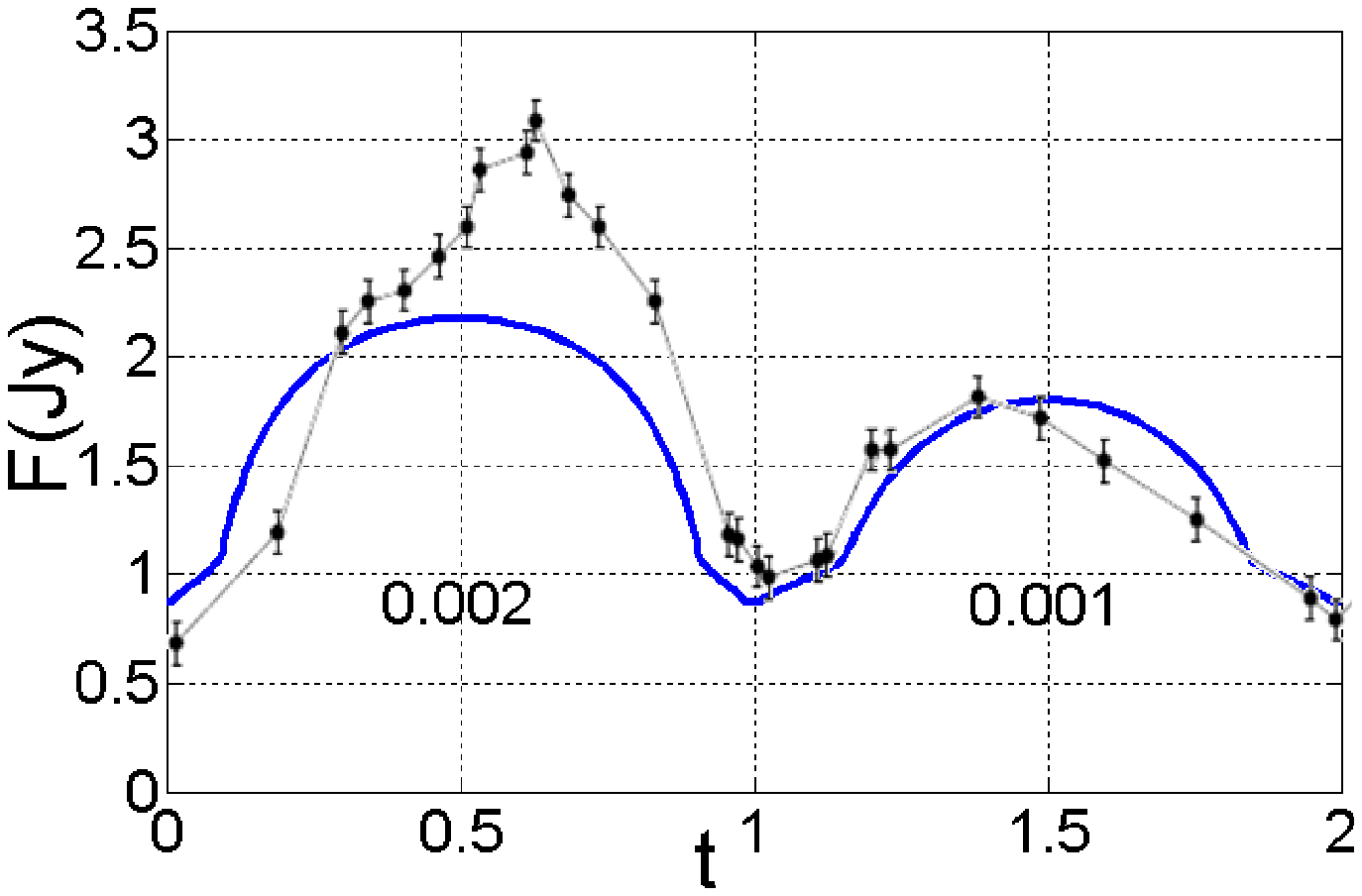}}
\vskip -7.2 cm
\caption{The calculated radio-flux light curve and the observed one
from White et al. (2005).
$F=L_{\rm radio}/ 4 \pi D^2$, where $L_{\rm radio} $ is given by
equation (\ref{lradio}) (with $k_r=0.07$), and $D=2.3 \kpc$ is the distance to $\eta$ Car.
In the 1992-1998 period we use our standard parameters, while in
the 1998-2003 cycle we take $\eta_B=0.001$ instead of $\eta_B=0.002$.
We don't try to fully fit the radio light curve, but just to
demonstrate that the parameters can fit the observation, and
emphasize the role of the weak magnetic field in the primary wind.
}
\label{compare}
\end{figure}

{{{ To further illustrate the effect of some variables, in Figure \ref{compare2}
we repeat the presentation of Figure \ref{compare} but we changed the ionizing flux from
$\dot N_{i2}=2\times 10^{48} \s^{-1} {\rm sr}^{-1}$
to   $\dot N_{i2}=3.5 \times 10^{48} \s^{-1} {\rm sr}^{-1}$,
and the efficiency parameter from $k_r=0.07$ to $k_r=0.04$.
The values of $\eta_B$ are given on the lines. }}}
\begin{figure}
\resizebox{0.99\textwidth}{!}{\includegraphics{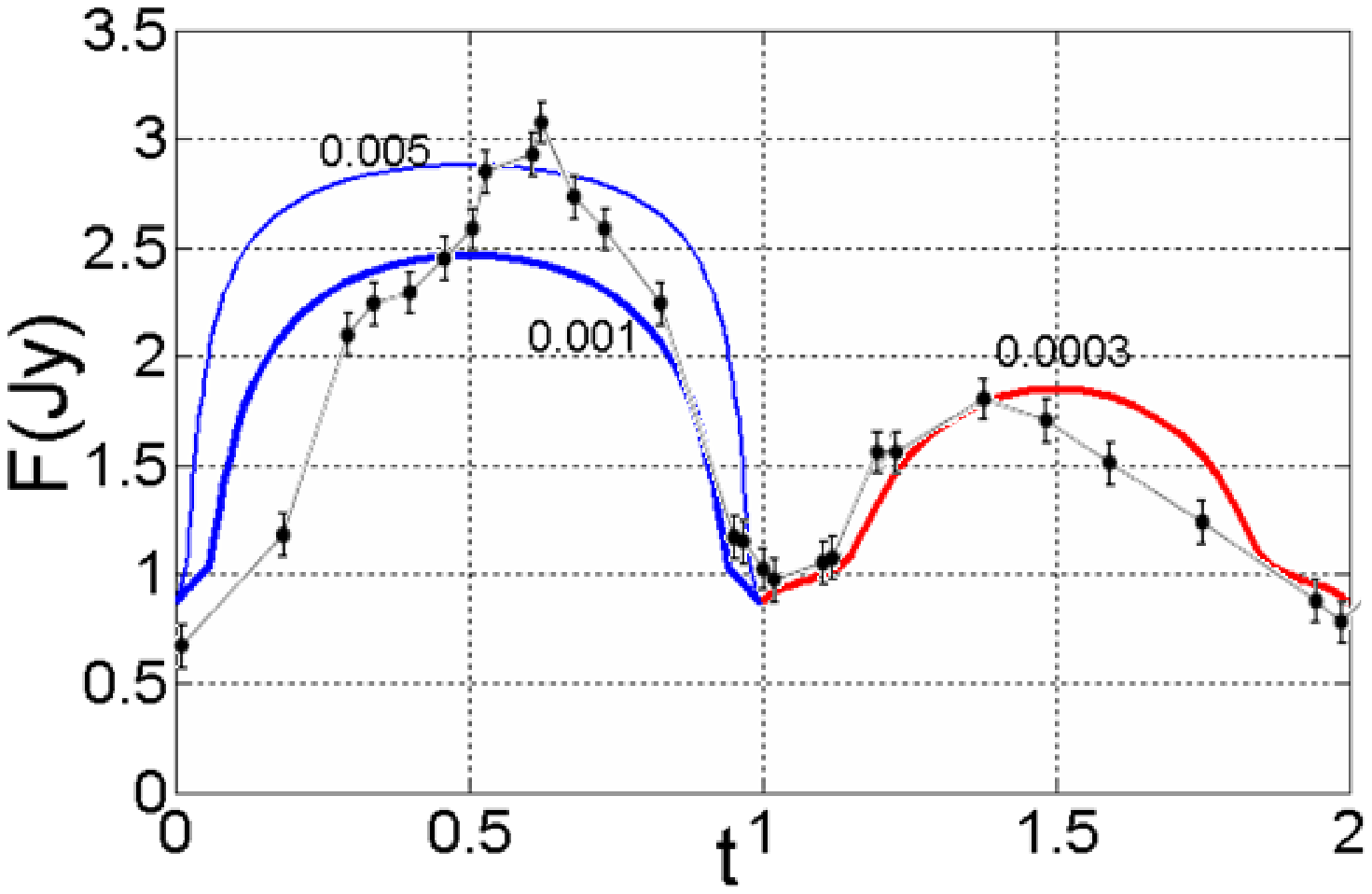}}
\vskip -7.2 cm
\caption{Like Figure \ref{compare}, but the ionizing flux is increased by a
factor of 1.75, and the efficiency of producing radio emission reduced from
$k_r=0.07$ to $k_r=0.04$. The value of $\eta_B$ is given for each curve.}
\label{compare2}
\end{figure}

\section{FAST VARIATIONS NEAR PERIASTRON PASSAGE}
\label{peri}

The accretion process near periastron passage might lead to the formation of
an accretion disk for a very short time (Soker 2003, 2005a), and might even
lead to a transient launching of two opposite jets, as suggested theoretically
(Soker 2005a; Akashi et al. 2006), and might have been observed (Behar et al. 2006).
Basically, in a steady state Bondi-Hoyle type accretion
flow the accreted mass has not enough specific angular momentum to form an accretion disk.
However, stochastically accreted blobs at the onset of the accretion phase
might lead to the formation of a transient accretion disk, and possibly to two jets.

Because the $3.5 \cm$ observations (White et al. 2005) don't sample the time near periastron,
we consider the 7~mm and 1.3~mm light curves reported by Abraham et al. (2005).
In Figure \ref{xraymm} we compare the temporal evolution of the X-ray flux (Corcoran 2005)
with the 7~mm flux and the 1.3~mm flux from Abraham et al. (2005) near periastron.
Zero time is taken at June 29, 2003.
The scales and vertical displacements are not the same for the different graphs, as we
are only interested in the temporal behavior.
In Figure \ref{xraymm}, the 1.3~mm flux varies between 5~Jy and 23~Jy, and the
7~mm flux varies between 0.7~Jy and 3.4~Jy.
The 1.3~mm flux is much larger that the 7~mm flux as a result of contribution from hot
dust to the 1.3~mm flux (Cox et al. 1995), and because optical depth is lower at 1.3~mm.
\begin{figure}
\resizebox{0.89\textwidth}{!}{\includegraphics{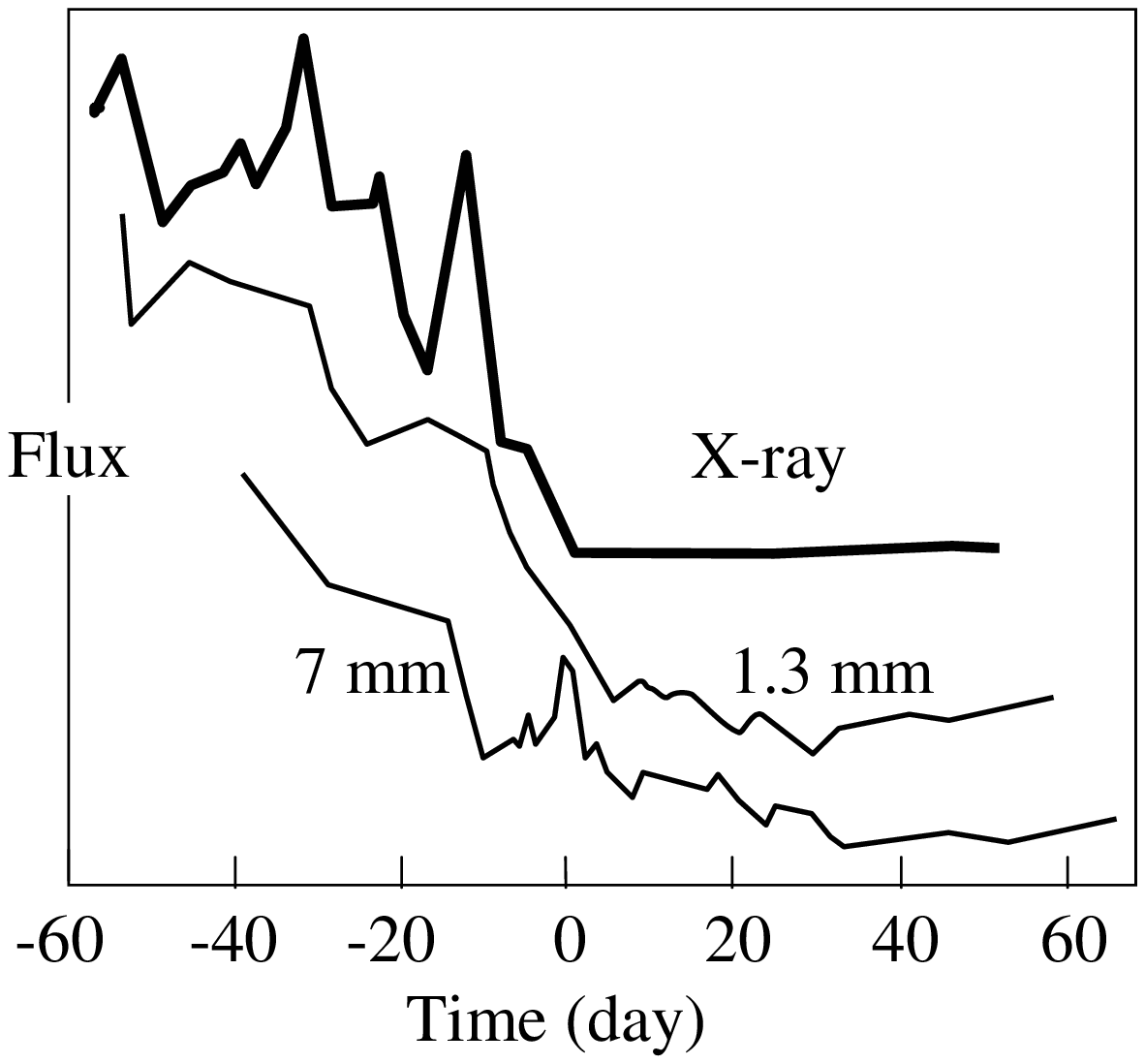}}
\vskip -8.2 cm
\caption{X-ray (Corcoran 2005)  and millimeter (Abraham et al. 2005) emission
from $\eta$ Car at the beginning of the spectroscopic event,
drawn to emphasize the peaks and deeps in the emission curve.
The three graphs don't have the same flux scale and are shifted vertically arbitrarily.
Zero time is at June 29, 2003. }
\label{xraymm}
\end{figure}

There is some correlation between the X-ray and millimeter emission behavior.
This can be understood as follows. From Figure \ref{absf} we learn that near
periastron all ionizing radiation escape from the
opening formed by the secondary wind, i.e., with in angle $\alpha \la 60^\circ$
(but not during the accretion phase).
We ignored in our treatment the fast orbital motion near periastron that causes
this opening cone to wind around the secondary, therefore increasing absorption
(Nielsen et al. 2007).
This winding of the winds-interaction region is not enough to account
for the X-ray decline, and for that the accretion model was proposed
(Soker 2005b; Akashi et al. 2006).
The accretion process is likely to be stochastic, especially at the beginning.
At the beginning we expect blobs of material to be accreted (Soker 2005b).
This accretion is assumed to shut down the secondary wind.
The stochastic accretion at the beginning of the accretion phase is likely
to lead to a few weeks of sporadic ejection of the secondary wind, mainly
perpendicular to the orbital plane.
The sporadic fast ejecta from the secondary is the explanation of the high X-ray peaks
observed just before the long X-ray minimum (Soker 2005b; Akashi et al. 2006).

The sporadic ejection of the secondary wind will increase the radio emission as well.
When dense blobs are accreted they will absorb more of the secondary radiation
simultaneously with the reduction in the secondary wind and X-ray emission.
When the secondary wind is intensified and lead to X-ray flare, it also cleans a
region around the primary, allowing more ionizing radiation to escape.
This will increase the radio emission.

The accreted matter might also form a blanket at a distance of
a few solar radii, or $\sim 0.1-0.3 R_2$, above the secondary (Soker 2007a) .
This reduces the effective temperature and as a result of that the
ionizing flux.
As evidence from Figure \ref{absf}, the ionizing rate is close to the
absorption rate. This implies that any small change in the ionizing flux
can lead to a large change in the difference between emission and absorption.
The last peak in 7~mm emission (a small knee is seen also in the 1.3~mm light curve
at that time) that corresponds to no X-ray flare, might be
a short heating episode of the secondary surface.

X-ray flares exist along the entire orbit, but they are much stronger near periastron
passages (Corcoran 2005). The X-ray variations along most of the orbit result from
stochastic variations in the winds properties and instabilities in the winds collision
process.
According to the accretion model (Soker 2005b; Akashi et al. 2006) the larger
variations near periastron passages result from instabilities in the accretion process.
We propose the same type of behavior for the radio emission.
Namely, large variations in the free-free radio emission near periastron passages
result from instabilities in the accretion process that cause large variations
in the escaping ionizing radiation.

\section{SUMMARY}
\label{summary}

Our main goal was to explore the effect of the interaction of the two winds in $\eta$ Car
binary system on the radio emission.
The general flow structure and ionization and recombination rates
we deal with are far too sensitive to the different parameters for us to
build an exact model.
Our results use, and support, previous claims that the $3.5 \cm$ radio emission from
$\eta$ Carinae can be explained by free-free (bremsstrahlung) emission from ionized gas
(Cox et al. 1995; Duncan \& White 2003), and the ionization source is mainly the binary
secondary star (Verner et al. 2005; Duncan \& White 2003; Soker 2007a).
What we have shown here is that the absorption of ionizing photons by the shocked
primary wind must be considered in addition to the absorption by the freely flowing wind
(Soker 2007a).

The radio emission is likely to come from dense regions at distances of
hundreds to thousands AU from the binary system. Therefore, the observed radio flux
depends sensitively on the unknown distribution of mass in these region.
We parameterized the radio emission efficiency of the ionized gas
by a parameter $k_r$. We found that $k_r \simeq 0.07$ in our phenomenological model.

{{{
The optical depth of the primary wind to $3 \cm$ radio emission at distances smaller than
$\sim 300 \AU$ from the system is large (eq. \ref{tau1}), and most radio emission
comes from larger distances.
The optical depth is smaller at smaller wavelengths. Therefore, the exact radio light curve
will be different at different wavelengths.
This different behavior at different wavelengths ($2,~6,~20 \cm$) is seen in the very
eccentric massive binary star system WR~140 (HD193793; White \& Becker 1995).
However, the mass loss rates in WR~140 are much lower than those in $\eta$ Car, and no
direct comparison can be made.
}}}

The ionizing photon rate reaching large distances from the secondary is the difference
between the emitted rate by the secondary star (Verner et al. 2005; Duncan \& White 2003;
Soker 2007a), with some contribution from the primary, and the absorption rate.
The influence of the ionizing photon rate is explored in Figure \ref{radiof1}.

The absorption rate is sensitive to the primary wind mass loss rate and
velocity, and to the compression of gas in the shock.
As the primary stellar wind collides with the secondary wind it passes through
a shock wave, cools to low temperatures ($\sim 10^4 \K$), and is compressed by the
ram pressure of the winds until its magnetic pressure equals the ram pressure.
Our main new finding is that the compressed post-shock primary wind has a crucial
influence on the absorption rate, hence on the ionization state of the circumbinary
material, and consequently on the observed radio flux.
The absorption rate by the freely-flowing primary wind and by the shocked
primary wind is presented in Figure \ref{absf}.
The magnetic pressure determines the compression factor of the postshock primary wind.
This is parameterized by $\eta_B$ (eq. \ref{etaB}), which is the ratio of the preshock
magnetic pressure to the ram pressure of the primary wind.
The influence of $\eta_B$ on the observed radio flux is demonstrated in Figure \ref{compare}.

The role of the magnetic fields in the primary wind might be far beyond determining the
compression factor.
Soker (2007b) speculated that the 20 years long Great Eruption of $\eta$ Car that occurred
160 years ago was triggered by magnetic activity in the inner convective region
of $\eta$ Car.
This activity expelled the outer radiative zone which is now the bipolar nebulae,
the Homunculus.

The winds properties determine also the opening angle of the low-density (shock) cone,
$\phi_a$ (Figute \ref{fwinds2}).
Close to periastron passage most, or even all, of the ionizing photons
escape through the low-density cone (again, for $\sim 10$~weeks very close to periastron
passage the low-density cone does not exist in our model because of the accretion process).
Therefore, the opening angle of the cone
is another source of sensitivity of the radio emission on the winds properties.

In section 4 we commented on the fast variation in the
7~mm and 1.3~mm emission reported by Abraham et al. (2005).
We accept the proposed accretion model for the spectroscopic event
(Soker 2005b; Akashi et al.\ 2006; Soker \& Behar 2006; Soker 2007a).
In that model for $\sim 10$~week near periastron passage the secondary
star accretes mass from the primary wind instead of blowing its own wind.
The accretion phase implies the following.
(1) There is no low-density cone any more.
(2) A high density accretion column (see figure \ref{fwinds1}) absorbs all
radiation to the direction opposite that of the primary ($\alpha \sim 0$);
(3) The accreted mass can form a blanket around the secondary, reducing its
effective temperature and the ionizing flux from its surface (Soker 2007a).
(4) At the onset of the accretion phase the accretion process can be stochastic
and the accretion material posses a relatively large specific angular momentum.
This process can lead to the sporadic launching of a collimated secondary wind.

The sporadic launching of the secondary wind can lead to X-ray flares (peaks
in X-ray emission), and reduces absorption of ionizing radiation by clearing a path
through the dense primary wind. This is our explanation to a possible correlation between
X-ray peaks and millimeter emission peaks (figure \ref{xraymm}).
The next spectroscopic event is predicted to occur in January 2009.
We predict that near that time, i.e., fall of 2008 to January 2009,
there will be some correlations between the X-ray and radio intensity.
We also predict that signatures of collimated outflow might be detected
during these `flares'.
For example, strong blue-shifted X-ray lines (Behar et al. 2006).

Some processes were not included in our calculation of the observed radio flux.
Future more sophisticated models will have to include these effects.
In particular we did not consider the following processes.
\begin{enumerate}
\item We did not consider the winding (wrapping) of the low-density cone around the
binary system near periastron passage (Nielsen et al. 2007).
This process affects results near periastron passage, but not too close, when
our model assumes that the low-density cone does not exist because of the accretion phase.
For most of the rest of the time the winding has a small effect.
\item
The contribution of the primary star to the ionizing flux was not considered here.
\item
The dependence of the primary wind properties on latitude (Smith et al. 2003)
was ignored in our treatment.
\item
The long recombination time at large distances (eq. \ref{trec}), implies that in a
more sophisticated model the response time of the ionized gas to the ionizing
radiation should also be considered.
\end{enumerate}

{{{ We thank Diego Falceta Goncalves and an anonymous referee for useful comments. }}}
This research was supported by a grant from the
Asher Space Research Institute at the Technion.

\end{document}